\documentclass[twocolumn,dvipsnames,nofootinbib,floatfix,groupedaddress,superscriptaddress]{revtex4-1}

\usepackage{hyperref}
\PassOptionsToPackage{unicode}{hyperref}
\PassOptionsToPackage{bookmarks=false}{hyperref}

\hypersetup{colorlinks=true, linkcolor=blue, urlcolor=blue, citecolor=blue}

\usepackage{braket}
\usepackage{amsmath}
\usepackage[utf8]{inputenc}
\usepackage{bbold}
\usepackage{geometry}
\usepackage{fullpage}
\usepackage{graphicx}
\usepackage{paralist}
\usepackage{subfigure}
\usepackage{slashed}
\usepackage{float}
\usepackage{color,xcolor}
\usepackage{units}
\usepackage{mathrsfs}
\usepackage{enumitem}
\usepackage{bm,paralist,xspace,url}

\graphicspath{{./PlotsPRL/}}

\newcommand{\cdf}{\mathop{\mathrm{cdf}}}

\usepackage[margin=false,inline=true,index=true,draft]{fixme}
\fxsetup{
  inlineface=\footnotesize
}
\fxusetheme{colorsig}

\FXRegisterAuthor{oleg}{anoleg}{O.R.}
\FXRegisterAuthor{alex}{analex}{A.M.}
\FXRegisterAuthor{kirill}{ankirill}{K.B.}

\begin{document}
\title{New physics at the Intensity Frontier:  how much can we learn and how?}

\author{Oleksii~Mikulenko}
\email{mikulenko@lorentz.leidenuniv.nl}
\affiliation{Instituut-Lorentz, Leiden University, Niels Bohrweg 2, 2333 CA Leiden, The Netherlands}
\author{Kyrylo~Bondarenko}
\affiliation{Nordita, KTH Royal Institute of Technology and Stockholm University, Hannes Alfv\'ens v\"ag 12, 10691
Stockholm, Sweden}
\author{Alexey~Boyarsky}
\affiliation{Instituut-Lorentz, Leiden University, Niels Bohrweg 2, 2333 CA Leiden, The Netherlands}
\author{Oleg~Ruchayskiy}
\affiliation{Niels Bohr Institute, University of Copenhagen, Blegdamsvej 17, DK-2010, Copenhagen, Denmark}

\date{1 December 2023}

\begin{abstract} 
Intensity Frontier experiments are often evaluated by the smallest coupling it can probe, irrespective of what particle can be found or the scientific significance of its detection. In this work we propose a new framework that determines the number of events required to characterize new particle properties. For example, we show that Heavy Neutral Leptons require 100 events to establish the neutrino mass hierarchy, and 1000 events to reveal the Majorana phase of active neutrinos. Ultimately, this framework presents a more objective way to connect experiments to their scientific outcomes.

\end{abstract}
\maketitle

The Standard Model of Particle Physics (SM) has achieved remarkable success but falls short in providing a complete explanation for all observable phenomena in the universe. Several unresolved questions remain, including the nature of dark matter, the origin of neutrino masses, the mechanism 
matter-antimatter asymmetry of the Universe
generation, etc. This strongly suggests the existence of some new physics beyond the SM.

One of the possible solutions to the SM puzzles could be \textit{feebly interacting particles} (FIPs) \cite{Shaposhnikov:2007nj,Alekhin:2015byh,Lanfranchi:2020crw,Agrawal:2021dbo}. 
Such particles can be searched at colliders, but only for sufficiently large masses and short decay lengths~\cite{Bondarenko:2023fex}. Dedicated experiments are necessary to efficiently explore long-lived light FIPs, and this has influenced the further development of the Intensity Frontier (IF) program at CERN~\cite{Bonivento:2013jag,Alekhin:2015byh,SHiP:2015vad,Beacham:2019nyx,EuropeanStrategyforParticlePhysicsPreparatoryGroup:2019qin} and globally~\cite{Hewett:2014qja}. 

The assessment of the potential of the Intensity Frontier projects is often based on their capacity to exclude part
of the parameter space, if new particles \emph{are not discovered}, quantified by the 
\textit{minimal coupling constant} necessary for a detectable signal.
Yet, another important question is what we can learn if new particles \emph{are discovered}, to what extent do they help to address the challenges in the Standard Model? Although the answer depends on the specific problem, 
the potential of an experiment in this regard is
generally characterized by
the \textit{maximum signal count} that can be achieved within the space of unconstrained parameters.

In this Letter, we present an approach to  
assessing the potential of a given experiment
based on its ability to probe physically interesting properties of  Feebly Interacting Particles following their detection. We describe a systematic framework designed for this approach and,  as an example, apply it to the model of Heavy Neutral Leptons (HNLs), focusing on their role in explaining the origin of neutrino masses.

\textbf{Method.} In the following, we provide a brief summary of the proposed framework. 
The details and code~\cite{modeltesting} that implements such analyses are discussed in the companion paper~\cite{PRD}.

Consider the extension of the SM with a FIP. Such an extension incorporates: 
    \textbf{(i)} a coupling constant $\alpha \ll 1$, dictating the number of observable events in an experiment and 
    \textbf{(ii)} a set of couplings $g_a \sim 1$, shaping the model's internal structure and defining its phenomenology for a given $\alpha$.

For each point in the parameter space $g_a$ we aim to compute the number of events, $N_t(\alpha)$, sufficient to conclude if the corresponding model can explain one of the BSM phenomena (i.e., is consistent with a given set of external data
at a certain confidence level). 
This number is used to determine the value of the minimal coupling constant $\alpha$ for a given experiment.

The average number of events in each observable channel is 
\begin{equation}
    \lambda_i = N_\text{t}(\alpha) \times \mathrm{Br}_i  (g_a)\times \epsilon_{\text{det},i}  + b_i.
\end{equation} 
Here $\mathrm{Br}_i$ is the branching ratio of the $i$-th visible channel, $\epsilon_{\text{det},i}$ is the mean detection efficiency, and $b_i$ is the average background count.
Then, given a set of observed counts, $s_i$, we define the CL for the ``tested'' model, given the data, $ \text{CL} = 1- 
\mathscr{P}_A \times  [1-\cdf(\chi^2|\lambda)] $.
It takes into account the prior of the model $\mathscr{P}_A$ as defined by external experimental data as well as the probability of the data, $s_i$, given the model (where $\chi^2$ is defined in the usual way), see \cite{PRD} for details. A ``real'' model, which is the subject of the sensitivity estimate, leads to random counts $s_i$, which exclude the tested model at CL with some probability. 
By requiring that the probability exceeds some predefined threshold $P$, we can determine the number of events $N_\text{t}(\alpha)$ sufficient to claim that the model can/cannot explain the BSM phenomenon at the given confidence level.

Below, we provide a step-by-step example of the use of the method. 
Throughout the analysis, we choose the threshold values $\mathrm{CL} = 0.9$ and $P = 90\%$.

\textbf{Example: neutrino physics and HNL parameters.}
A possible solution to the origin of the neutrino masses is the Type I seesaw~\cite[Ch.~14]{ParticleDataGroup:2020ssz} --- an extension of the SM by $\mathcal N$  Heavy Neutral Leptons. 
This model possesses rich phenomenology and multi-dimensional parameter space, making it an excellent example of the approach described above.

Assuming the detection of a signal, consistent with the hypothesis of HNL, the natural question to be addressed is whether we found \textit{the explanation} of the neutrino masses and to what extent this explanation is complete. To be specific, we will focus on the analysis of two heavy neutral leptons, the minimal number capable of explaining the two mass differences observed in the neutrino oscillation data~\cite{ParticleDataGroup:2020ssz}.

\begin{figure*}[!t]
    \centering
    \includegraphics[height = 0.26\linewidth]{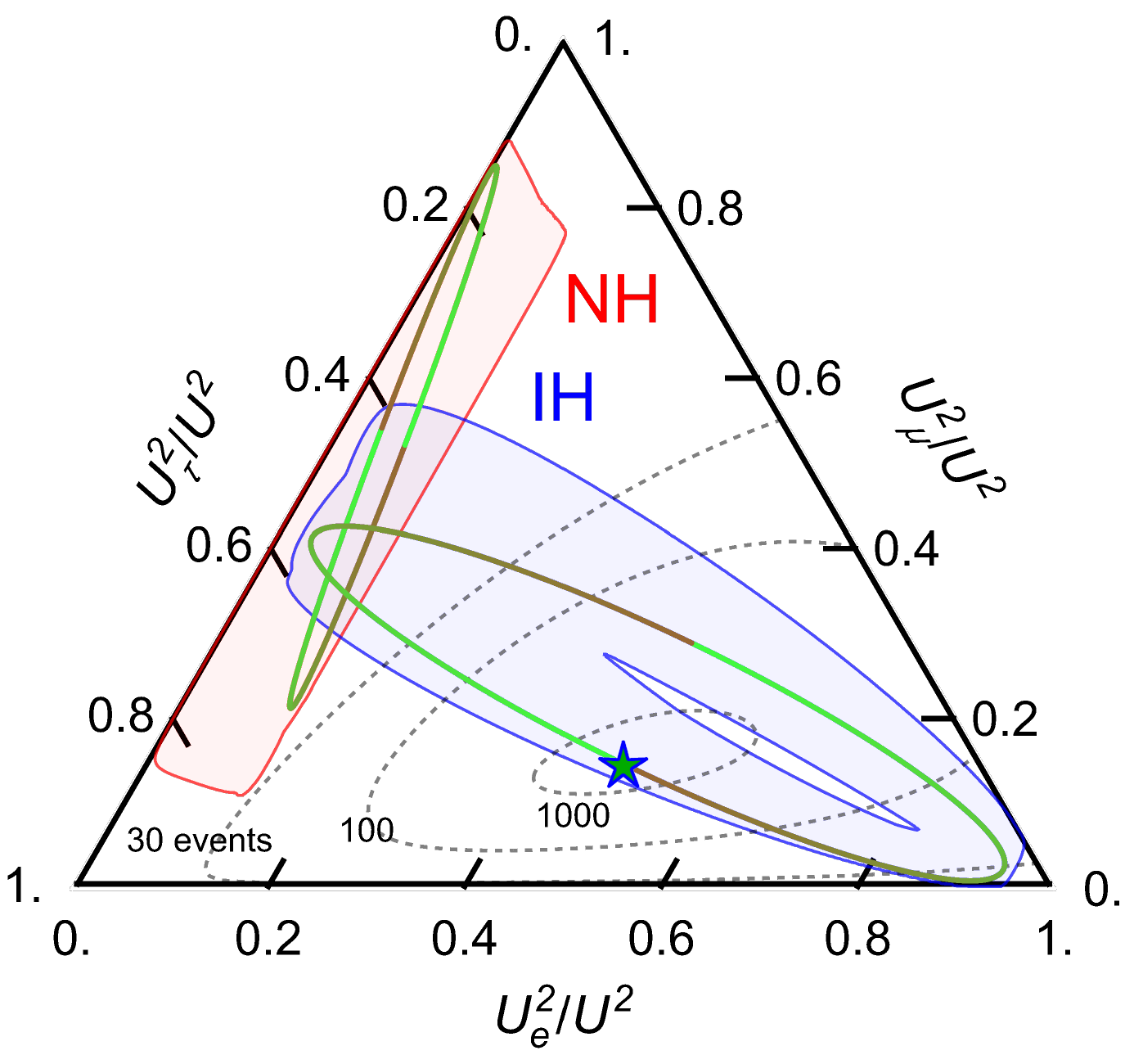}~\hfill~
    \includegraphics[height = 0.27\linewidth]{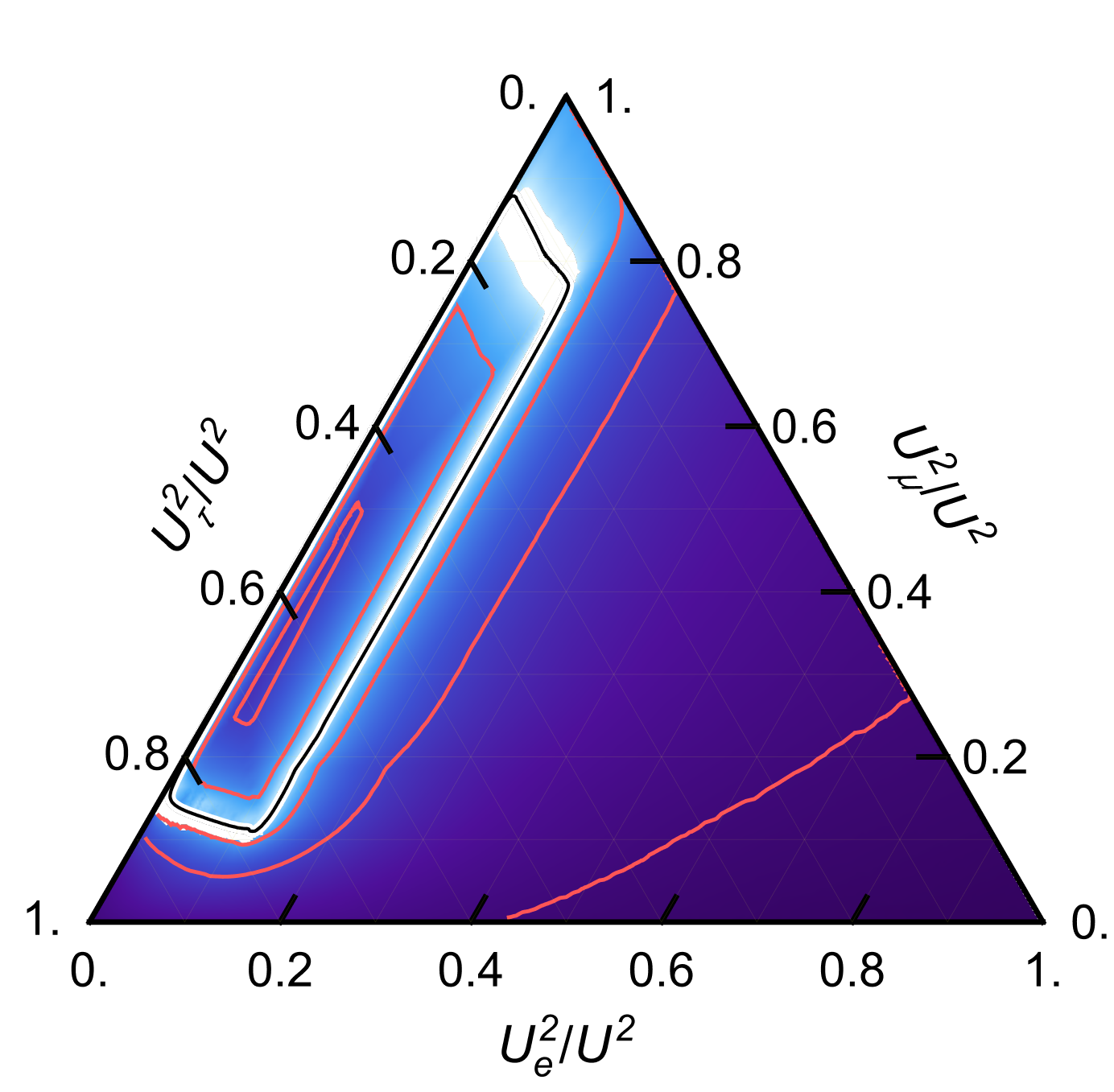}~\hspace{-4mm}~
    \includegraphics[height = 0.27\linewidth]{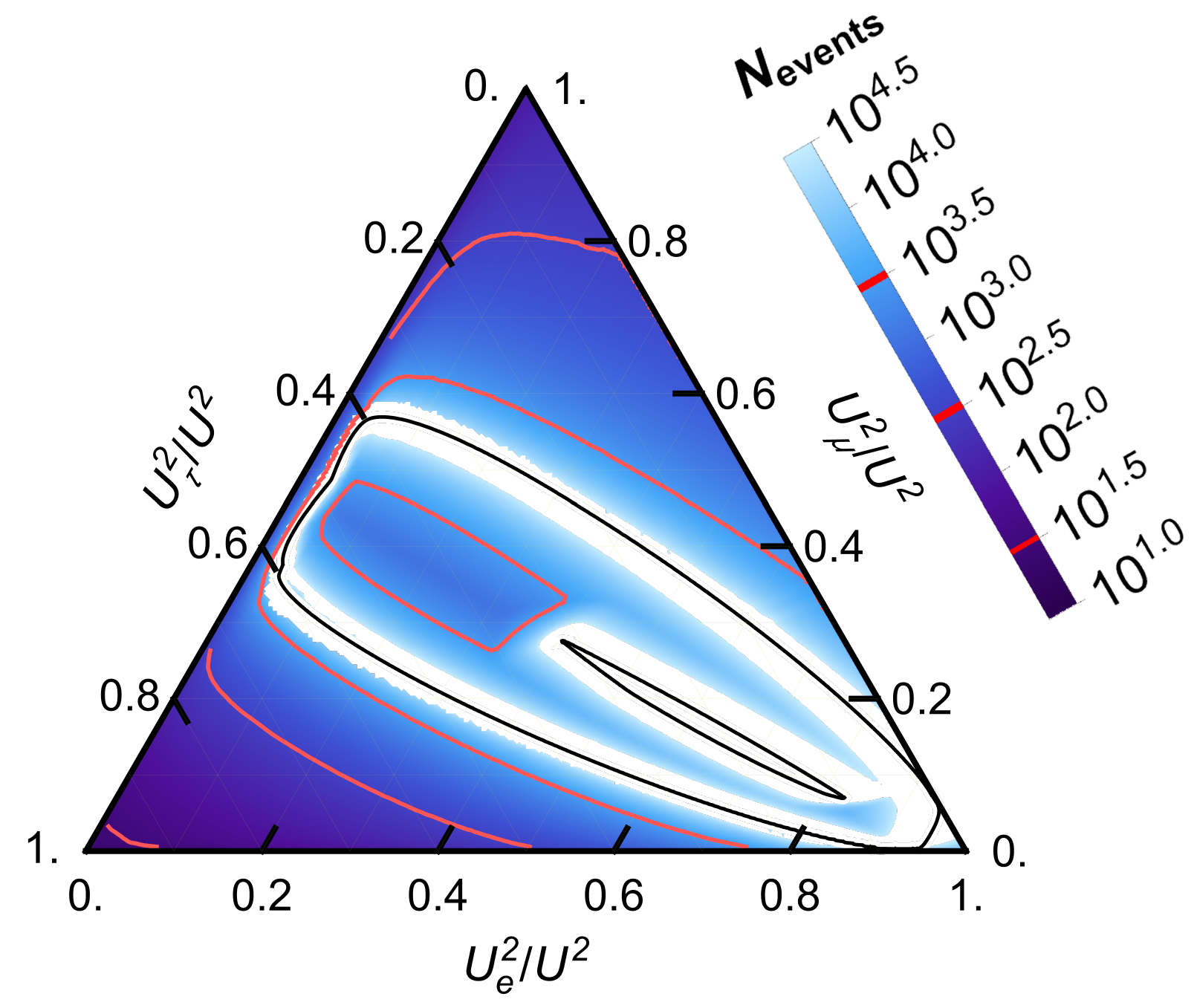}     
    \caption{\textbf{Left Panel:} parameter
    space, where two degenerate HNLs
    may generate neutrino masses with normal (red) and inverse (blue) hierarchy.
    The shaded areas align with the 90\% CL variations in neutrino oscillation parameters~\protect\cite{NuFIT5.2} The brown-green lines correspond to the best-fit values for each hierarchy, where only the neutrino Majorana phase $\eta$ varies from zero (brown) to $\pi$ (green). For each point on the ellipse, two phases $\eta$ and $\eta+\pi$ are possible. The dashed lines are reconstructed contours for 30, 100, and 1000 observed events for parameters of the benchmark model (star). 
    \textbf{Right Panels:} the required count of HNL decay events to evaluate the consistency or inconsistency of the measured mixing ratios $U^2_\alpha/U^2$ with the aforementioned HNL model and neutrino hierarchy as indicated by the black lines. Points closer to these lines require higher precision, thus increasing the required number of events. The red contours in the triangular graphs correspond to the red lines in the color legend. HNL mass is fixed at $m_N=\unit[1.5]{GeV}$.
    }
    \label{fig:illustration}
  \end{figure*}

The phenomenology of HNL is determined by four parameters: the mass $m_N$ of the detected HNL and three mixing angles $U_\alpha^2$ \cite{gorbunov:2007ak,Bondarenko:2018ptm}, reparametrized 
as the total mixing angle $U^2 = \sum_\alpha U^2_\alpha$ and \textit{mixing ratios} $x_\alpha \equiv U_\alpha^2/U^2$. 
If $U^2$ is sufficiently small to ensure that the particle decay length $l_\text{decay}\propto U^{-2}$ exceeds the experiment's length scale, the number of events scales as $N_\text{tot} \propto U^4$. 
The lower bound on the value of  $U^2$  is given by the seesaw bound: $U^2_\text{seesaw}\equiv \sum m_i/{m_N}$. This scale lies below the discovery reach of the future IF experiments~\cite{Abdullahi:2022jlv}. Therefore, two HNLs, if they are to be observed in the region $U^2\gg U^2_\text{seesaw}$, must form a quasi-Dirac pair,
such that their overwhelming contributions of order $m_N U^2$ cancel each other to neutrino masses $m_\nu$. In general, this requirement does not prevent the two species from having different masses. However, the case of two mass-degenerate HNL with $\Delta M\ll m_N$ is of particular interest. First, it realizes an approximate lepton symmetry, with two Majorana particles forming a quasi-Dirac fermion. Second, it implies oscillations in the sterile sector and has the potential to generate the observed baryon asymmetry of the Universe (see e.g. work~\cite{Klaric:2020lov} and refs therein). 
  
To verify that the particles discovered at an IF experiment explain neutrino masses, the following questions must be addressed:

\begin{enumerate}
    \item Are the observed phenomenological parameters $m_N$ and $U^2_\alpha$ consistent the seesaw mechanism? In the following analysis, we will focus on this question.
    \item Are there two Majorana HNLs? If two signals with different masses are observed, the necessary check is to verify that the required cancellation of their contributions to neutrino masses is possible $(m_{N} U^2_{\alpha})_1\approx(m_{N} U^2_{\alpha})_2$ with the measured parameters of each candidate. If two HNLs are mass-degenerate, they would create a single combined signal. Resolving the two species may be possible by observing the oscillations between them~\cite{Tastet:2019nqj}, which are most discernible at the mass difference $\Delta M \sim \unit[10^{-6}]{eV}$. Otherwise, a single new particle offers little information: It may be a Dirac fermion with zero contribution to neutrino masses or a Majorana particle, requiring further search for the missing ingredients.
    \item Ultimately, does the cancellation of two HNL contributions yield the proper neutrino masses? In this case, very precise measurements of the HNL parameters must be made, with uncertainty $U^2$ at the level of the seesaw bound. Such high precision is not achievable with the near-future Intensity Frontier experiments.
\end{enumerate}

In our analysis, we rely on the connection between the couplings of the HNL to different lepton flavors $U_\alpha$, and the properties of active neutrino oscillations. Neutrino oscillation data restrict the mixing ratios~\cite{Casas:2001sr} to specific regions of the ternary plot. Under the fine-tuning assumption $U^2\gg U^2_\text{seesaw}$, these regions are shown in Figure~\ref{fig:illustration}, cf.
also~\cite{Ruchayskiy:2011aa,Bondarenko:2021cpc,Tastet:2021vwp,Drewes:2022akb, Krasnov:2023jlt}.
The shape of these regions depends on the measured values of the neutrino parameters.
Notably, if all the known neutrino parameters are fixed, there remains a single unconstrained Majorana phase, $\eta$.\footnote{In the parametrization of the PDG~\cite{ParticleDataGroup:2020ssz} there are two phases $\eta_1$, $\eta_2$. However, when the lightest neutrino is massless, one phase can be absorbed by the redefinition of the neutrino fields.} As $\eta$ is changed from $0$ to $2\pi$, ellipses are drawn in the ternary plot.

\begin{figure*}[!t]
    \centering
\begin{minipage}{.7\textwidth}
\centering
        \includegraphics[width = 0.33\textwidth]{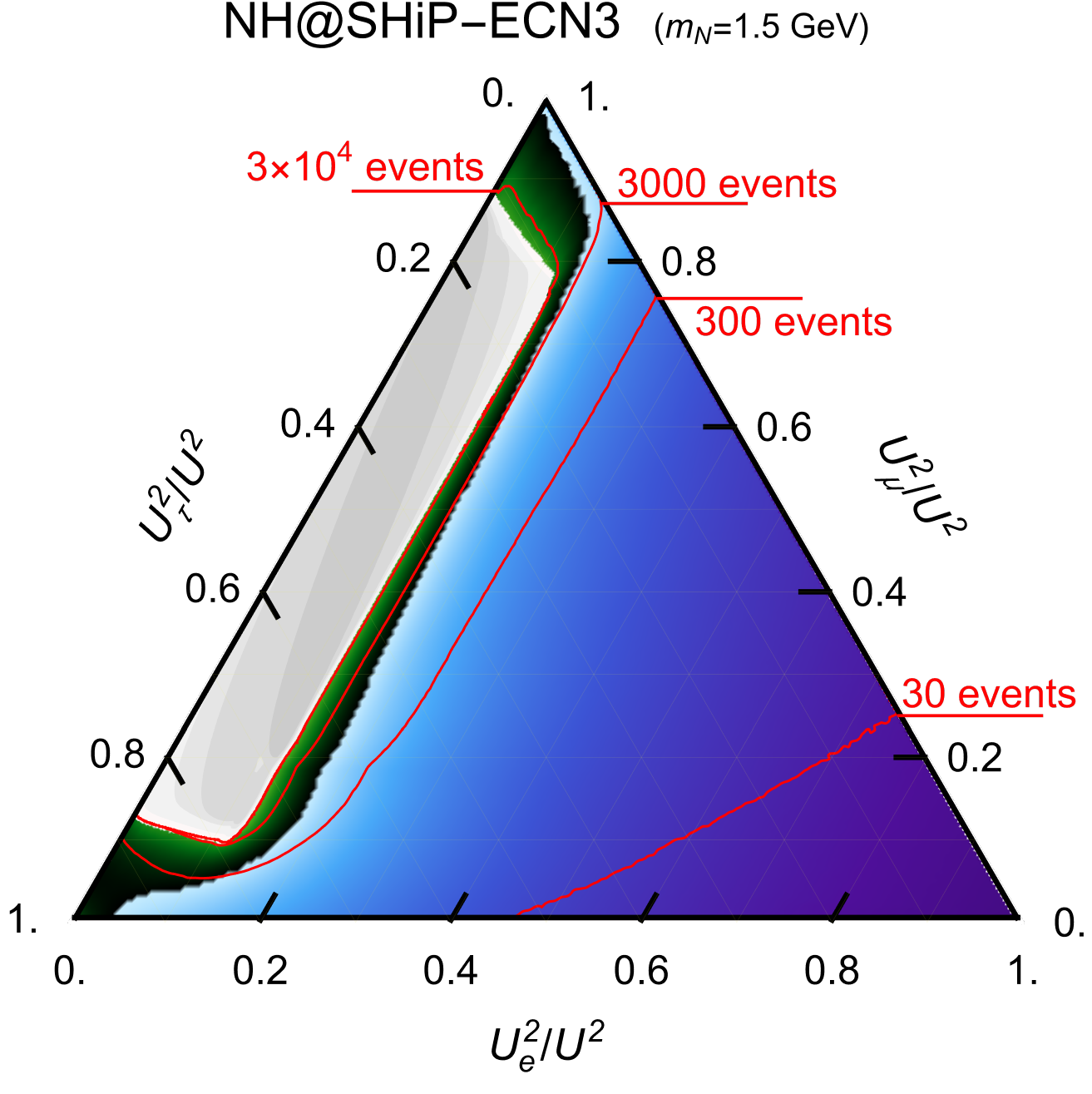}~
        \includegraphics[width = 0.33\textwidth]{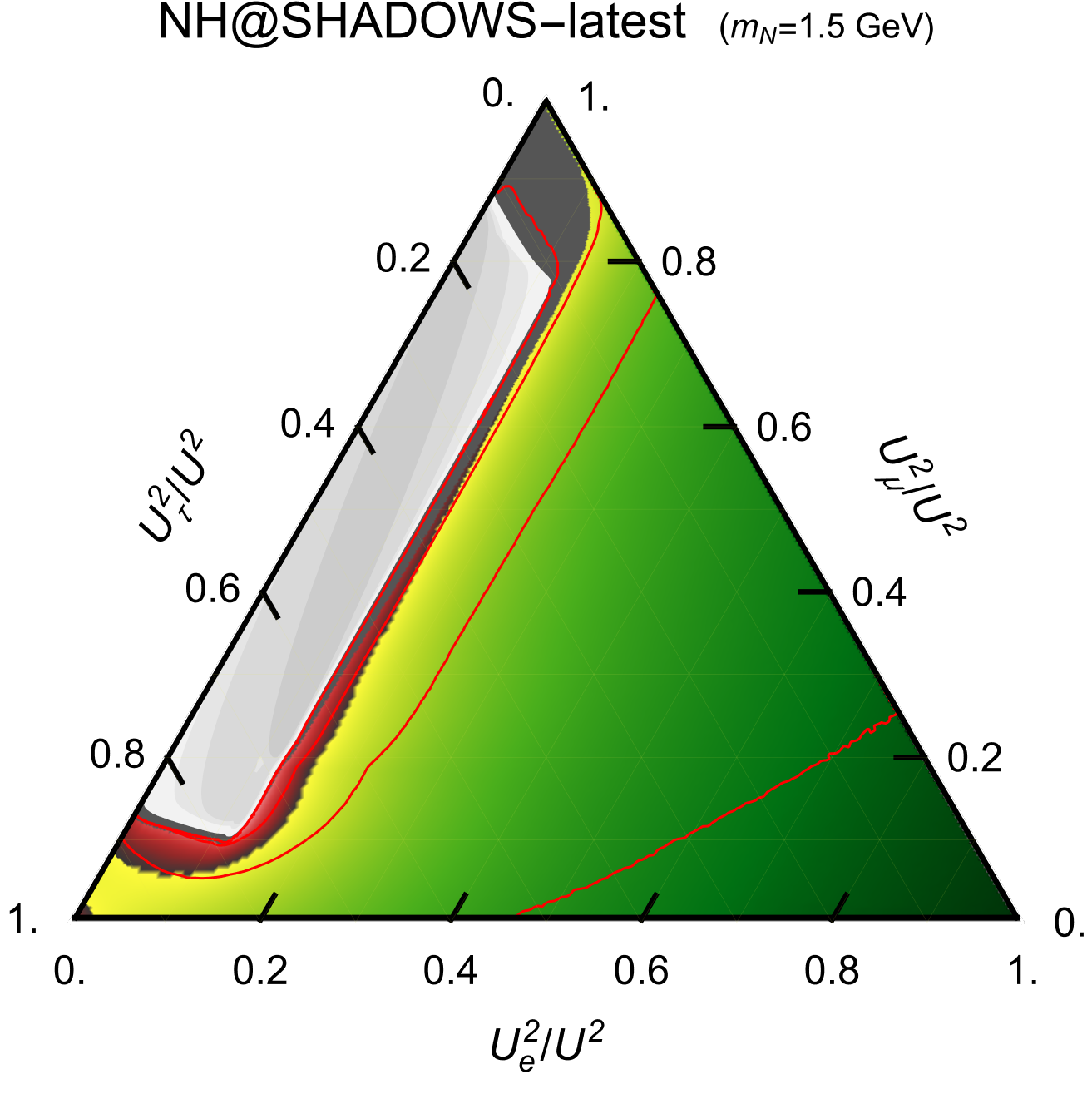}~
        \includegraphics[width = 0.33\textwidth]{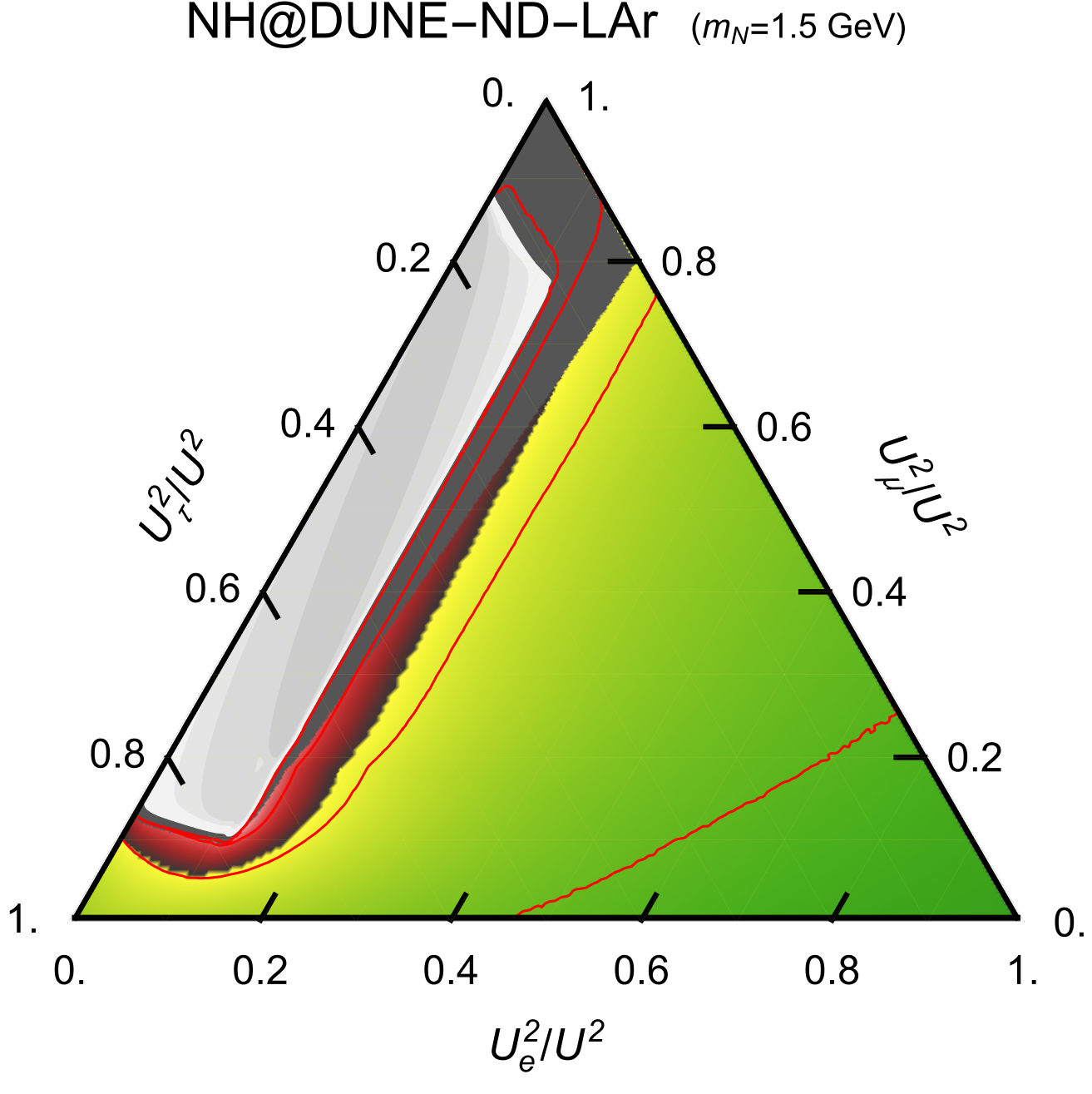}
        
        \includegraphics[width = 0.33\textwidth]{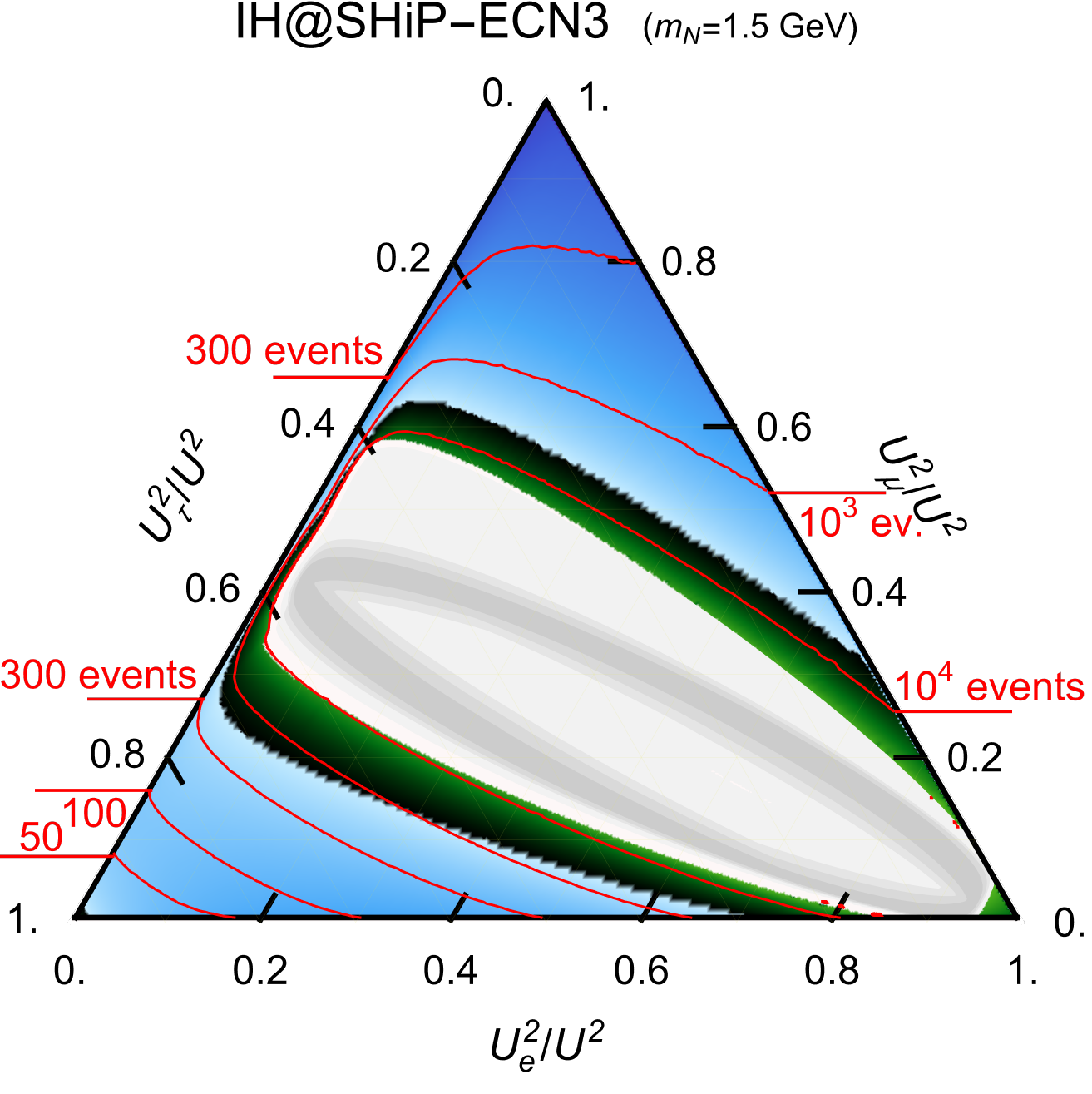}~
        \includegraphics[width = 0.33\textwidth]{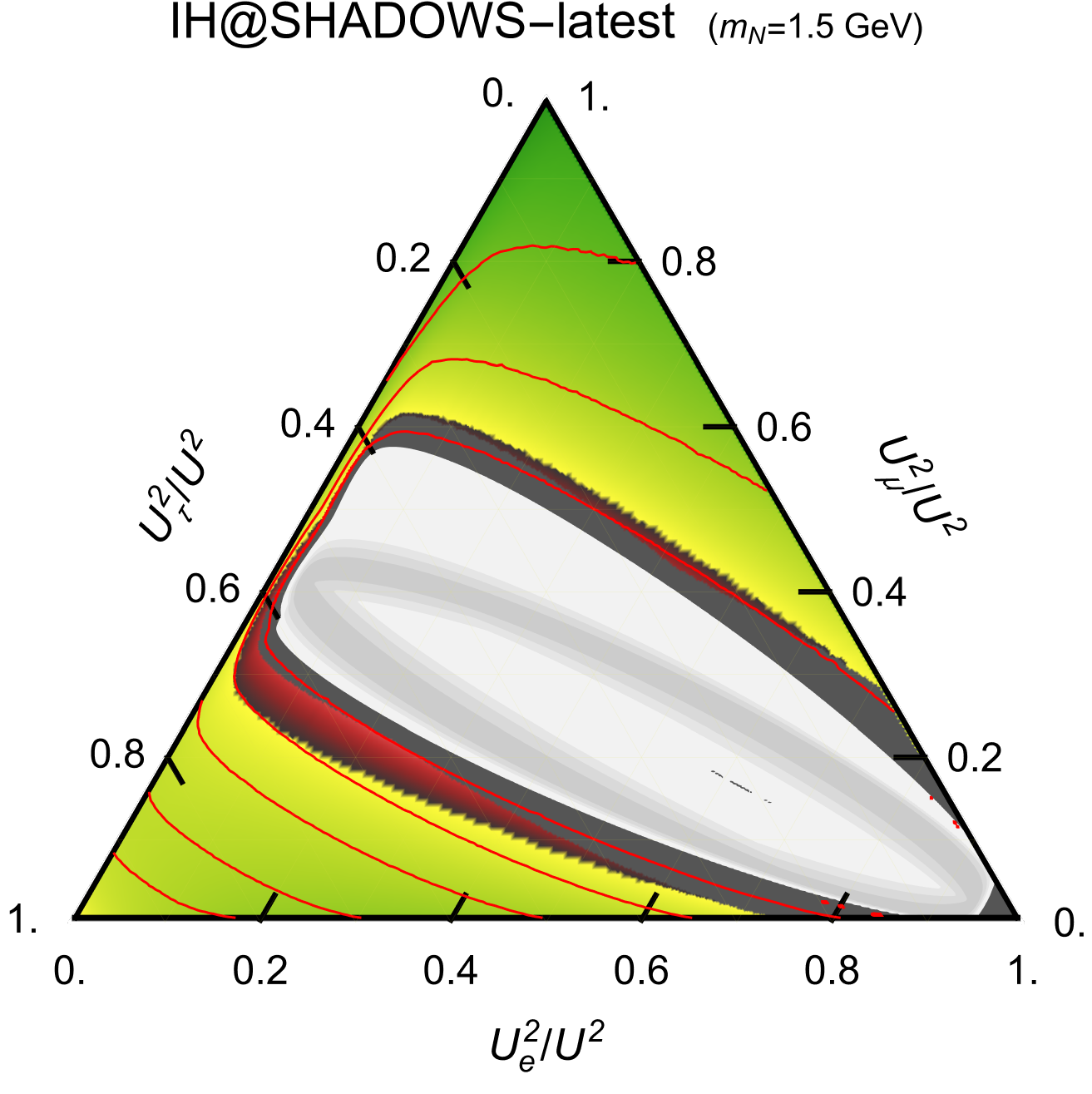}~
        \includegraphics[width = 0.33\textwidth]{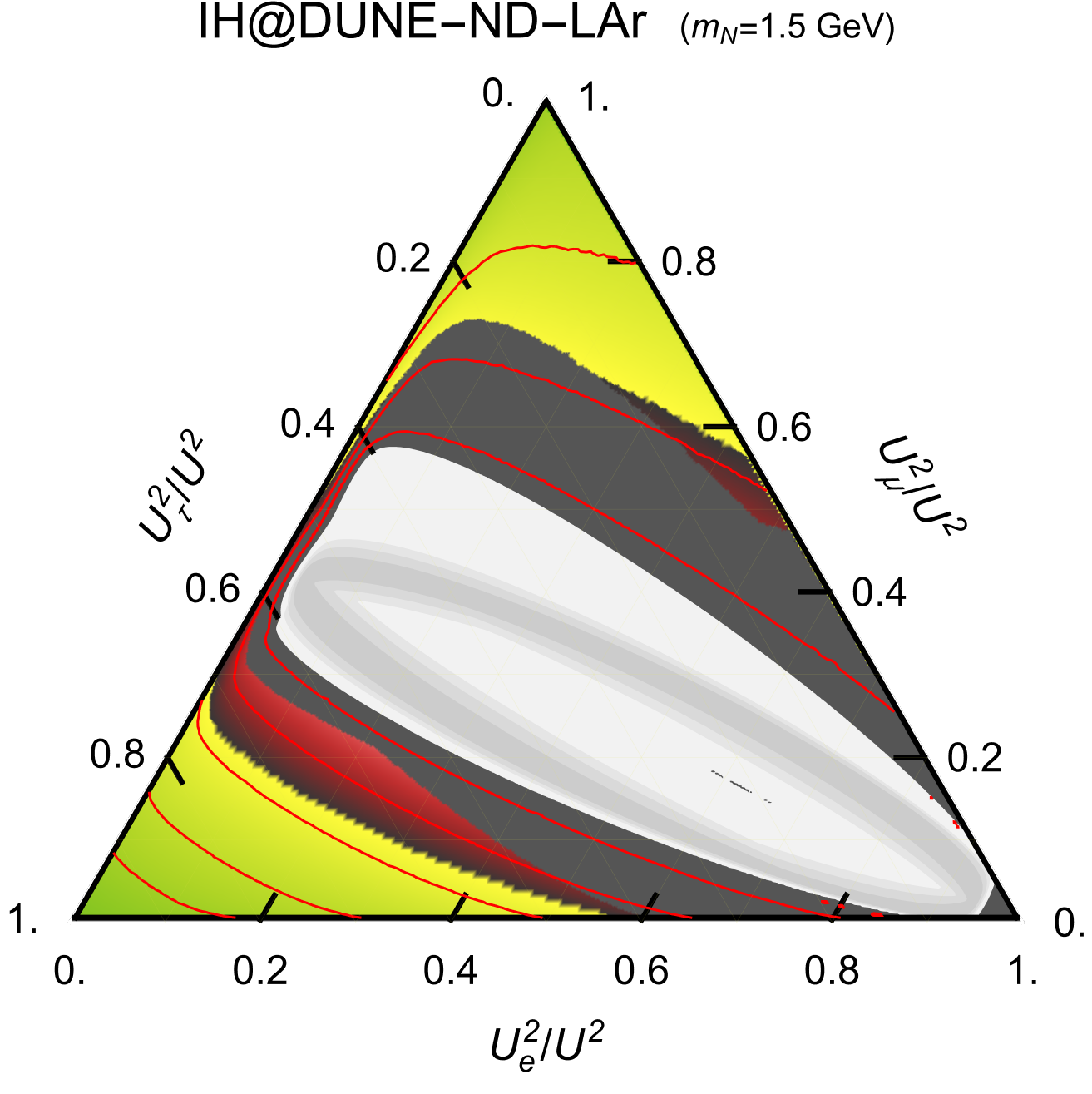}~
\end{minipage}~
\begin{minipage}{.15\textwidth}
\centering
    \includegraphics[height = 7cm]{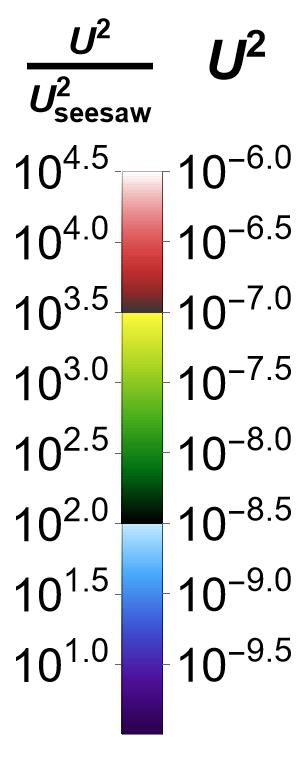}

\end{minipage}
    \caption{
    The limits on the total coupling constant $U^2$ to rule out specific neutrino mass hierarchies. \textit{Light-gray regions:} Parameter space aligned with the two HNL seesaw mechanism; darker shades indicate more favored values from neutrino oscillation data~\protect\cite{NuFIT5.2}. \textit{Colored regions:} The values of  $U^2$  needed to dismiss NH (top)/IH (bottom) in SHiP~\cite{SHiP_BDF_ECN3}, SHADOWS~\cite{SHADOWS_BDF}, and DUNE~\cite{DUNE:2020fgq}, both in absolute terms and normalized to $U^2_\text{seesaw}$. Three color schemes differentiate typical sensitivity for SHiP ($10^{-10}-10^{-8.5}$, blue) against SHADOWS and DUNE ($10^{-8.5}-10^{-7}$, green), and the $U^2$ range ($>10^{-7}$, red) which may be excluded by prior experiments for \textit{certain} $x_\alpha$ combinations~\cite{Bondarenko:2021cpc,Abdullahi:2022jlv}.
    The dark gray regions (confined within the red regions) are combinations of $U^2$, $x_\alpha$ that are already excluded by previous experimental searches. 
\textit{Red lines:} isocontours of the observed number of events, corresponding to the plotted $U^2$. The relation between the number of events and $U^2$ has been computed using the \textsc{SensCalc} package~\cite{Ovchynnikov:2023cry}.
}
    \label{fig:hierarchy_exclusion_maps}
\end{figure*}

For the HNL analysis, we use the results of~\cite{Bondarenko:2018ptm} and consider the following 6 channels: three leptonic ones ($ee$, $e\mu$, $\mu\mu$) and 3 semi-leptonic (\textsc{nc}, $eh$, and $\mu h$). 
Here, \textsc{nc} refers to HNL decays mediated by neutral current interaction with neutrinos and neutral mesons ($N \to h^0 \nu_\alpha$ with $h^0 = \pi^0, \eta, \eta',\rho^0$, \dots) in the final state, while $h$ denotes charged mesons $\pi^\pm, \rho^\pm, K^\pm, D^\pm, \dots$.
The first 4 channels contain an undetectable neutrino ($N \to \ell_\alpha \ell_\beta \nu_\beta$ and $N \to h^0\nu_\alpha$), while the last 2 channels allow us to reconstruct the HNL mass as the invariant mass of the lepton-hadron pair. Within each channel, the branching ratios for all decay modes exhibit dependence on $x_\alpha$. Analysis of the modes separately may only verify the \textit{consistency} of the observed data with the assumption of HNL observation.

With a sufficiently large number of detected events, we can discern whether the flavor ratios align with the normal or inverted mass ordering, or are inconsistent completely with the hypothesis of only two HNLs. Furthermore, the ratios $x_\alpha$ can be determined so precisely, that it can, in combination with better determination of neutrino parameters, determine the range of Majorana phases.

\begin{figure*}[t!]
    \centering
    \includegraphics[width = \linewidth]{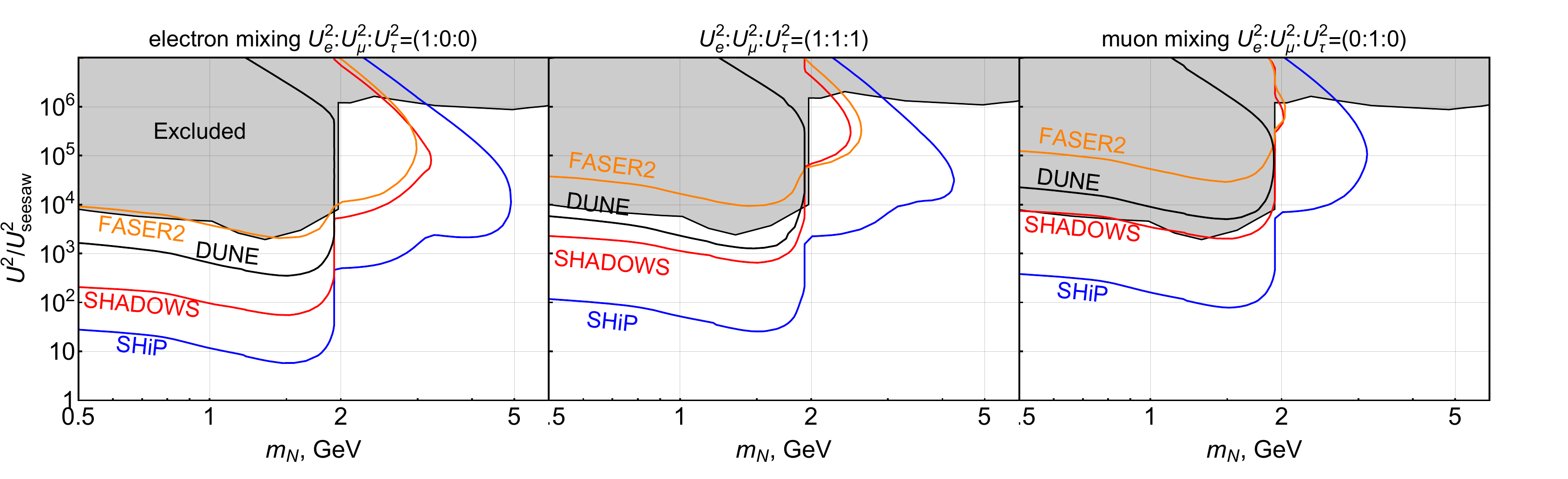}
    \caption{
    Sensitivity regions where the experiment can rule out the seesaw mechanism with two degenerate HNLs for specific mixing patterns, assuming a normal hierarchy. The mixing patterns (pure electron, equal couplings, or pure muon) are labeled at the top of each panel. The colored lines represent FASER2, DUNE, SHADOWS, and SHiP (from top to bottom).
      }
    \label{fig:exp_sensitivity}
\end{figure*}

\bigskip
\textbf{Results.}
With the methodology outlined above, we can now present the results achievable by an Intensity Frontier experiment, assuming a signal has been detected.

In our analysis, we consider an HNL with mass $\unit[1.5]{GeV}$. Furthermore, we assume the most optimistic scenario of an experiment with the highest detection efficiency $\epsilon_{\text{det},i} = 1$ for all visible channels and the background-free regime $b_i=0$.

Increasing the number of events reduces the uncertainty in the determination of $x_\alpha$ and eventually allows differentiation between mass orderings. For both hierarchies and each point $x_\alpha$, inconsistent with the considered hierarchy, we found the number of \textit{observed} events $N_\text{ev}(x_\alpha)$ required to claim inconsistency.

Figure~\ref{fig:hierarchy_exclusion_maps} shows the limits on the coupling constant $U^2$ required to reject the hypothesis of NH (top) or IH (bottom). The limits are shown for the future beam-dump experiments: SHiP~\cite{SHiP_BDF_ECN3}, SHADOWS~\cite{SHADOWS_BDF}, and DUNE~\cite{DUNE:2020fgq}. Note that, while $N_\text{ev}(x_\alpha)$ for the given hierarchy does not depend on the experiment, the limits on the coupling constant differ. To relate the coupling constant to the observed number of events, we used the \textsc{SensCalc} package~\cite{Ovchynnikov:2023cry}.

The SHiP experiment can exclude most of the ternary plot with a coupling constant as small as $U^2\sim 10^{-9.5}-10^{-8.5}$ for the HNL mass $\unit[1.5]{GeV}$. The SHADOWS experiment achieves a similar sensitivity in the range $U^2\sim 10^{-8}-10^{-7}$. In contrast, the DUNE experiment can put constraints only from $U^2\gtrsim 10^{-7.5}$ and has a relatively large parameter space in which the exclusion of the IH is not possible, given the current constraints on $U^2_\alpha$~\cite{Abdullahi:2022jlv}.

The number of required events depends on the ID capability of the experiment detector system. The determination of the ratio of $x_e$ and $x_\mu$ can be achieved through the analysis of $eh$ and $\mu h$ decays,
which are based on the identification of electrons, muons, and charged $\pi$, $K$ mesons. The \textsc{nc} channel decays, mostly resulting in neutral meson and photons in the final state, provide sensitivity to the total $U^2$ and, therefore, allow to exclude degeneracy along the line of constant $x_e/x_\mu$ ratio. Without insight from the \textsc{nc} decays, such as $N\to \nu \pi^0$, the total required number of HNL events $N_\text{tot}$ can increase by up to a factor of ten.

\bigskip
\textbf{Conclusion.}
The sensitivity of Intensity Frontier experiments is typically gauged by their capacity to observe a few events to claim a discovery of new physics. In case of such a breakthrough, the examination of the \textit{properties} of the new particle would be the main goal of further research, requiring observation of a substantially larger signal yield. The forthcoming Intensity Frontier experiments have the ability to collect an ample signal within previously uncharted parameter spaces.

This study delves into the potential learnings about the beyond the Standard Model phenomena in the event of a discovery at an Intensity Frontier experiment. To this end, we develop a comprehensive framework that provides the estimate for the signal yield necessary to establish a relation between the observed new physics and the BSM problems. In light of numerous experimental proposals, this opens up a completely new way of determining the sensitivity of an experiment to new physics. We exemplify the use of the framework through the study of Heavy Neutral Leptons, a model capable of explaining neutrino oscillations.

In the minimal extension of the SM with two Heavy Neutral Leptons that generate active neutrino masses, the interactions of the new particles are constrained by the neutrino oscillation data. Our findings indicate that, in some parts of the parameter space, only $\gtrsim 100$ events would suffice to measure the coupling constants precisely enough to claim (in-)consistency with the minimal scenario. Furthermore, under the assumption of the minimal model, insights into the properties of active neutrinos may be gained, such as the hierarchy of the neutrino masses. With $\gtrsim 1000$ events, a determination of the active neutrino Majorana phase becomes possible.

The best potential for inferring the properties of new particles in the GeV mass range is offered by the SHiP experiment. The maximum number of events detectable at SHiP over 15 years of operation can reach $\mathcal{O}(10^6)$. Such a large amount offers a rich resource for detailed examination of new physics as illustrated in this Letter. In addition, it has the potential to resolve HNL oscillations, driven by their mass splitting~\cite{Tastet:2019nqj}. In this case, SHiP can not only determine the parameters responsible for neutrino oscillations but also explore the parameter space of leptogenesis in the $\nu$MSM-like model~\cite{Asaka:2005an,Asaka:2011pb} (leptogenesis through oscillations, driven by two HNL). Thus, the SHiP experiment provides a robust and comprehensive platform for the identification, detection, and understanding of potential new physics.

The method, presented here, can be employed for other feebly interacting models, including not only portals but also models augmented by higher-dimensional operators~\cite{DeVries:2020jbs,Liang:2023yta}.
This underscores the robustness and adaptability of our approach, ensuring its relevance and applicability across a broad spectrum of theoretical frameworks in particle physics.

\bibliographystyle{apsrev} 
\bibliography{bibliography}

\end{document}